# Mapping the Dirac fermions in intrinsic antiferromagnetic topological insulators $(MnBi_2Te_4)(Bi_2Te_3)_n$ (n=0, 1)

Zuowei Liang[1], Mengzhu Shi[1], Qiang Zhang[1], Simin Nie[2], J.-J. Ying[1], J.-F. He[1], Tao Wu[1], Zhijun Wang[3,4], Zhenyu Wang[1*] and X.-H. Chen[1*]

[1]Department of Physics and Chinese Academy of Sciences Key laboratory of Strongly-coupled Quantum Matter Physics, University of Science and Technology of China, Hefei, Anhui 230026, China
[2]Department of Materials Science and Engineering, Stanford University, Stanford, California 94305, USA
[3]Beijing National Laboratory for Condensed Matter Physics, and Institute of Physics, Chinese Academy of Sciences, Beijing 100190, China
[4]University of Chinese Academy of Sciences, Beijing 100049, China

*Correspondence and requests for materials should be addressed to Z.W. (zywang2@ustc.edu.cn) and X.-H.C. (chenxh@ustc.edu.cn).

**Abstract**

Topological surface states with intrinsic magnetic ordering in the $MnBi_2Te_4(Bi_2Te_3)_n$ compounds have been predicted to host rich topological phenomena including quantized anomalous Hall effect and axion insulator state. Here we use scanning tunneling microscopy to image the surface Dirac fermions in $MnBi_2Te_4$ and $MnBi_4Te_7$. We have determined the energy dispersion and helical spin texture of the surface states through quasiparticle interference patterns far above Dirac energy, which confirms its topological nature. Approaching the Dirac point, the native defects in the $MnBi_2Te_4$ septuple layer give rise to resonance states which extend spatially and potentially hinder the detection of a mass gap in the spectra. Our results demonstrate that regulating defects is essential to realize exotic topological states at higher temperatures in these compounds.

The interplay between non-trivial band topology and magnetism provides a fertile ground for the realization of exotic quantum phenomena including the quantum anomalous Hall effect, axion insulator states, and Chiral Majorana modes (1-6). To achieve these states, one key step is to open a mass gap on the initial massless Dirac spectrum protected by the time-reversal symmetry. Currently, the prevalent approach to open this gap is either by magnetic doping (7,8) or through proximity-heterostructure engineering (9), where the material choice and property optimizations are extremely challenging. As of yet, only few examples of aforementioned phenomena have been experimentally achieved (3,4), let alone ready for practical applications. Therefore, an alternate arena for building magnetic topological materials is highly desired.

Recently, several numbers of $MnBi_2Te_4(Bi_2Te_3)_n$ compound series have been proposed as promising magnetic topological insulators with intrinsic A-type antiferromagnetic order (10-19). The magnetic moments, ferromagnetically ordered in a Mn sub-plane aligning with the c-axis (16), are predicted to break the time-reversal symmetry ($\Theta$) and thus open a sizeable mass gap on the (0001) cleaving surface. Soon after their discovery, quantized anomalous Hall conductivity (20) and quantum phase transition from axion insulator to Chern insulator (21) has been observed in moderate magnetic field. However, the detailed band structure of the topological surface states remains unclear, and photoemission data show controversial results. Unexpected gapless surface states have been recently reported by high resolution ARPES (22–26), in striking contrast with the large gap found in previous measurements (27, 28). Recalling the fact that local perturbations, such as defects, can give rise to localized resonances and nanoscale fluctuations of the topological surface states (29–32), it is thus critical to study Dirac electrons with a local probe.

Spectroscopic imaging scanning tunneling microscopy (SI-STM) offers a powerful tool to study the scattering properties and spatial inhomogeneity of the surface Dirac electrons. In this work, we use SI-STM to directly image the electronic properties of the topological surface states in $MnBi_2Te_4$ (Mn124) and $MnBi_4Te_7$ (Mn147). The quasiparticle interference patterns show clear evidence for the absence of backscattering revealing the spin-momentum locking of the surface states. We also find one type of defect that dominates in the $MnBi_2Te_4$ septuple layers (MBT SLs) but is absent in the $Bi_2Te_3$ quintuple layer (QL). These defects induce resonance states at or near the Dirac energy within a length scale larger than a few nanometers, which potentially impedes the detection of a gap in the tunneling spectra.

Single crystals of Mn124 and Mn147 were grown via a solid-state reaction method as described in earlier studies (11, 33). X-ray diffraction measurements were carried out on each sample to confirm its correct phase and high crystalline quality (Supplementary information part 1). The single crystals were cleaved at room temperature in a vacuum better than $1*10^{-10}$ Torr and immediately inserted into the STM head which remains at 5.5K. PtIr tips were used and their quality was tested on the surface of single crystal Au(111) before performing the measurements. Spectroscopic data were acquired by the standard lock-in technique at a frequency of 987.5Hz, under modulation voltage ~2-12mV.

The material with which we first focus on, Mn124, is a van der Waals (vdW) material formed by stacking Te-Bi-Te-Mn-Te-Bi-Te SLs (Fig. 1A). The crystals cleave naturally between two adjacent layers, resulting in a Te-terminated surface. The topographic image shows step-edge with a height of about 13.8 Å, consistent with the height of one SL layer (Fig. 1B). Figures 1d-f depict atomically resolved images, representing a few typical features. Topography at high bias voltage (-1V; Fig. 1D) clearly shows the hexagonal Te lattice with a periodicity of ~0.43 Å. Within this field of view, dark triangular defects are observed, corresponding to $Mn_{Bi}$ anti-site defects in the Bi layer beneath (32). The estimated spatial concentration of these defects is approximately 0.15 $nm^{-2}$, giving a value of 2.5% Mn/Bi substitution. The bright protrusions labeled as 'B' are most likely attributable to the Bi replacing the top layer Se atoms (34), which are rare on the surface. Upon approaching low energies, however, we find a third type of defect (type C) that dominates the entire surface (Fig. 1E and F; more bias-dependent topographies can be found in SI part 2). A closer look reveals that these features are weakly ordered somewhere, with their positions unchanged at different bias voltages. This indicates that these features are representative of the electronic structure response to strong local perturbations in the crystal that can be imaged at the surface. Unfortunately, these defects overlap with each other due to their large density, making it difficult to identify its species. (See SI part 3 for the positions and statistics of all three kinds of defects).

To better understand these defects, we have imaged the surface of Mn147, which has $Bi_2Te_3$ QLs intercalated between the nearest MBT SLs (Fig. 1B). Cleaving could expose two kinds of surfaces, either $Bi_2Te_3$ or MBT terminations. As shown in Fig. 1G, these two kinds of surfaces can be identified easily via their respective step heights, which in turn confirms the good quality of our crystals. Again, type C defects are clearly visible on the MBT SL surface but absent on the $Bi_2Te_3$ terminated surface, suggesting that these defects are intimately tied to the MBT layers. Our observations together with other recent reports of non-stoichiometric composition in both Mn124 ($Mn_{0.85}Bi_{2.1}Te_4$, ref.14, 11) and Mn147 ($Mn_{0.85}Bi_{4.1}Te_7$, ref. 19), support that type C defects are consistent with defects in the Mn layers, i. e., $Bi_{Mn}$ anti-site defects and Mn vacancies.

Our next step is to explore the topological surface states. The dI/dV spectra, which measure the local DOS against energy (in eV), are displayed in Fig. 1J for these three kinds of surfaces. The suppression of DOS in the energy ranging from -0.2 to -0.4eV can be attributed to the bulk gap that embraces the surface states, which will be discussed in greater detail later. Here we note that the spectra taken on the MBT SL in 124 and 147 show a distinct line shape, indicating that the electronic structure is quite sensitive to the stacking layers underneath (26). To obtain more information of the surface states, we employ Fourier-transform quasiparticle interference imaging (FT-QPI). In FT-QPI the standing wave pattern arising from the coherent scattering of quasiparticles can be measured as dI/dV (r, eV) maps and then Fourier-transformed to extract the scattering vectors which connect two states on the constant energy cuts. This is a powerful technique to measure the band dispersion and investigate the orbital/spin textures of the electronic states. To remove the set-point effect, we apply the Feenstra technique to normalize the conductance map before the Fourier-transform (35-36). Figures 2a-e show the resultant conductance maps at various energies obtained in Mn124 and the standing wave patterns are visible at high energies. The QPI patterns, as shown in Fig.2F-J, exhibit six-fold symmetric peaks along the ΓM direction.

To understand the momentum-space origin of these peaks, it is necessary to compare our QPI data with ARPES dispersion and theoretical calculations. It has been well established in $Bi_2Te_3$ and $Bi_2Se_3$ that the hexagonal warped shape of the surface band cut dominates the scattering channels far above the Dirac energy (37,38). Following this idea, we show the constant energy contour in the inset of Fig. 3a and label the q vectors along two high symmetric directions ΓM and ΓK as $q_1$-$q_6$. Indeed, the warping effect in Mn124 can be determined by the anisotropic dispersion along ΓM ($q_1$) and ΓK ($q_4$) extracted from the laser-APRES data (22,26) and our band structure calculations (SI part 4). On this basis, we can extract the dispersion of all possible scatterings (solid lines in Fig. 3a). A direct comparison with our STM data allows us to identify the QPI peak in our experiment as scattering vector $q_2 = \sqrt{3}k_{\Gamma K}$, a channel allowed by the helical spin texture (Fig.3B). The absence of $q_5$ along the ΓK direction in our QPI pattern indicates that backscattering is strictly forbidden far above the Dirac energy. At low energy where the dispersion is conic, the helical spin texture also forbids direct backscattering, thus attenuating the total QPI signal (Fig. 3C and Fig. 2J). Quantitative analysis of $q_2$ enables us to extract the dispersion of the Dirac fermions along the ΓK direction, and the results are summarized in Fig. 3D together with a typical dI/dV spectrum. The dispersion extracted from both QPI and ARPES data, and the minimum in the DOS, are all consistent with a Dirac point at -280±10meV (here we ignore the potential mass gap). Interestingly, the tunneling spectra also reveal double- peak features around -200meV, which corresponds to the splitting of bulk conductance band (CB1) in the AFM phase (22). We have also observed similar QPI patterns on both $Bi_2Te_3$ and MBT surfaces in Mn147 (SI part 5).

Having determined the topological nature of the surface state, we now discuss whether a Dirac-mass gap is opened or not by the bulk A-type AFM order. As shown in Fig. 3D, a finite DOS has been clearly observed near the Dirac point in Mn124, in good agreement with gapless spectrum found by high-resolution ARPES data (22-26). On the other hand, we do find a large amount of defects (2.5% of type A and about 10% of type C) in the samples. Since these defects could affect the local DOS and generate resonance states near the Dirac energy (32), it is critical to measure the spatial evolution of the DOS. To proceed, we acquire dI/dV (r, eV) spectra on a densely-spaced grid in a 15nm*15nm area of Mn124 (Fig. 4A). The differential conductance recorded near the Dirac point never reaches zero (Fig. 4B), in sharp contrast to an opening of a gap. However, the DOS is quite inhomogeneous, and seems to be directly related to the presence of type C impurities (see SI part 6). To obtain a visual sense of the variations of the spectra, we sort them into 20 bins against the dI/dV values at -281meV and then average all spectra in each bin (Fig. 4c; see SI part 7 for more details). This treatment well captures the inhomogeneity: At energies far away from the Dirac point, the spectral line shapes are quite similar, while near the Dirac point electronic resonance states emerge due to the presence of defects (see the inset for the difference between these spectra). Figure 4D maps the spatial distribution of the DOS along the arrow shown in Fig. 4B, which passes through the region showing minimum differential conductance in this field of view. The V-shaped bottom of this curve reveals that the decay length of the resonant state is larger than the average distance between the defects. Similar behavior can be found in Mn147, where sharp resonance peaks are clearly observed near Dirac energy with spectral weight spreading over distances of few nanometers (Fig. 4E-H).

The experimental realization of QAHE in Mn124 (20), which confirms the existence of a gapped Dirac state, and the finite DOS at the Dirac energy have led to an apparent paradox. The possible resolutions proposed so far include different spin reconstructions near the surface (22, 24, 26), the presence of magnetic domains (22) and weak hybridizations between Dirac fermions and the magnetic order (23). Our observations provide another possible explanation for the absence of mass gap in spectroscopic measurements. The extended states (~5nm) induced by defects are significantly overlapping in real space, and thus create a non-negligible spectral weight over the entire surface. These localized states are always generated in the vicinity of the Dirac energy impeding the detection of the mass gap of the mobile Dirac spectrum, especially when the gap is relatively small. Moreover, the energy position of the differential conductance minimum also varies within a range of 12meV, as shown in Fig. 4C. The existence of defect states and electronic disorder may also broaden the lineshape of photoemission spectra and thus smear a small gap in ARPES measurements. In fact, the coexistence of gap opening and gap filling owing to the magnetic dopants has been reported in V-doped $Sb_2Te_3$, where both full QAHE and a finite DOS near Dirac energy are observed (39, 40). To further isolate the effect of impurities from the surface spin reconstruction, dI/dV measurements under a magnetic field are required.

In summary, we have studied the scattering properties and spatial structures of the topological states in the newly discovered Mn124 and Mn147. Our QPI measurements reveal the helical spin texture of the surface states, confirming their topological nature. The spatial evolution of the local DOS allows us to identify the extended resonance states associated with native defects in MBT SL, suggesting that the finite DOS near Dirac energy may not be inconsistent with a mobility gap opening. These findings have established Mn124 and 147 as excellent platforms to realize exotic quantum phenomena, in addition to a path to obtaining ideal samples for the realization of these phenomena at higher temperatures.

**Acknowledgements**

We thank Vidya Madhavan, Lin Jiao, Shichao Yan and Jorge Olivares Rodriguez for helpful discussions.

**References**

1. Y Tokura, K Yasuda and A Tsukazaki. Magnetic topological insulators. *Nature Reviews Physics* 1, 126-143 (2019).
2. R Yu et al. Quantized anomalous Hall effect in magnetic topological insulators. *Science* 329, 61-64 (2010).
3. C-Z Chang et al. Experimental observation of the quantum anomalous Hall effect in a magnetic topological insulator. *Science* 340, 167-170 (2013).
4. D Xiao et al. Realization of the axion insulator state in quantum anomalous Hall sandwich heterostructures. *Physical review letters* 120, 056801 (2018).
5. M Mogi et al. magnetic heterostructure of topological insulators as a candidate for an axion insulator. *Nature materials* 16, 516 (2017).
6. QL He et al. Chiral Majorana fermion modes in a quantum anomalous Hall insulator–superconductor structure. *Science* 357, 294-299 (2017).
7. Q Liu, C-X Liu, C Xu, X-L Qi, S-C Zhang. Magnetic impurities on the surface of a topological insulator. *Physical review letters* 102, 156603 (2009).
8. YL Chen et al. Experimental realization of a three-dimensional topological insulator, $Bi_2Te_3$. *science* 325, 178-181 (2009).
9. F Katmis et al. A high-temperature ferromagnetic topological insulating phase by proximity coupling. *Nature* 533, 513 (2016).
10. J Li et al. Intrinsic magnetic topological insulators in van der Waals layered MnBi2Te4-family materials. *Science Advances* 5, eaaw5685 (2019).
11. J Cui et al. Transport properties of thin flakes of the antiferromagnetic topological insulator $MnBi_2Te_4$. *Physical Review B* 99, 155125 (2019).
12. Y Gong et al. Experimental realization of an intrinsic magnetic topological insulator. *Chinese Physics Letters* 36, 076801 (2019).
13. J-Q Yan et al. Crystal growth and magnetic structure of $MnBi_2Te_4$. *Physical Review Materials* 3, 064202 (2019).
14. A. Zeugner et al. Chemical Aspects of the Antiferromagnetic Topological Insulator $MnBi_2Te_4$. arXiv: 1812.03106 (2018).
15. B Chen et al. Intrinsic magnetic topological insulator phases in the Sb doped $MnBi_2Te_4$ bulks and thin flakes. *Nature communications* 10, 4469 (2019).
16. S. H. Lee et al. Spin scattering and noncollinear spin structure-induced intrinsic anomalous Hall effect in antiferromagnetic topological insulator $MnBi_2Te_4$. arXiv: 1812.00339 (2018).

17. C Hu et al. A van der Waals antiferromagnetic topological insulator with weak interlayer magnetic coupling. *arXiv:1905.02154* (2019).
18. J Wu et al. Natural van der Waals heterostructural single crystals with both magnetic and topological properties. *Science advances* 5, eaax9989 (2019).
19. RC Vidal et al. Topological electronic structure and intrinsic magnetization in $MnBi_4Te_7$: a $Bi_2Te_3$-derivative with a periodic Mn sublattice. *arXiv:1906.08394* (2019).
20. Y Deng et al. Magnetic-field-induced quantized anomalous Hall effect in intrinsic magnetic topological insulator $MnBi_2Te_4$. *arXiv:1904.11468* (2019).
21. C Liu et al. Quantum phase transition from axion insulator to Chern insulator in $MnBi_2Te_4$. *arXiv:1905.00715* (2019).
22. Y Chen et al. Topological Electronic Structure and Its Temperature Evolution in Antiferromagnetic Topological Insulator $MnBi_2Te_4$. *Physical Review X* 9, 041040 (2019).
23. H Li et al. Dirac Surface States in Intrinsic Magnetic Topological Insulators $EuSn_2As_2$ and $MnBi_{2n}Te_{3n+1}$. *Physical Review X* 9: 041039 (2019).
24. Y-J Hao et al. Gapless Surface Dirac Cone in Antiferromagnetic Topological Insulator $MnBi_2Te_4$. *Physical Review X* 9: 041038 (2019).
25. P Swatek et al. Gapless Dirac surface states in the antiferromagnetic topological insulator $MnBi_2Te_4$. *arXiv:1907.09596* (2019).
26. Y Hu et al. Universal gapless Dirac cone and tunable topological states in $(MnBi_2Te_4)_m (Bi_2Te_3)_n$ heterostructures. *arXiv:1910.11323* (2019).
27. M Otrokov et al. Prediction and observation of an antiferromagnetic topological insulator. *Nature* 576, 146 (2019).
28. E D L Rienks et al. Large magnetic map at the Dirac point in $Bi_2Te_3/MnBi_2Te_4$ heterostrctures. *Nature* 576, 423 (2019).
29. H Beidenkopf et al. Spatial fluctuations of helical Dirac fermions on the surface of topological insulators. *Nature Physics* 7, 939 (2011).
30. Z Alpichshev et al. STM imaging of impurity resonances on $Bi_2Se_3$. *Physical review letters* 108, 206402 (2012).
31. I Lee et al. Imaging Dirac-mass disorder from magnetic dopant atoms in the ferromagnetic topological insulator $Cr_x(Bi_{0.1}Sb_{0.9})_{2-x}Te_3$. *Proceedings of the National Academy of Sciences* 112, 1316-1321 (2015).
32. R R Biswas and AV Balatsky. Impurity-induced states on the surface of three-dimensional topological insulators. *Physical Review B* 81, 233405 (2010).
33. M Shi et al. Magnetic and transport properties in the magnetic topological insulators $MnBi_2Te_4(Bi_2Te_3)_n$(n= 1, 2). *Physical Review B* 100, 155144 (2019).
34. Y Jiang et al. Fermi-level tuning of epitaxial $Sb_2Te_3$ thin films on graphene by regulating intrinsic defects and substrate transfer doping. *Physical review letters* 108, 066809 (2012).
35. Rá Feenstra, JA Stroscio and Aá Fein. Tunneling spectroscopy of the Si (111) 2× 1 surface. *Surface science* 181: 295-306, 1987.
36. M Allan et al. Imaging Cooper pairing of heavy fermions in $CeCoIn_5$. *Nature physics* 9, 468 (2013).
37. L. Fu Hexagonal warping effects in the surface states of the topological insulator $Bi_2Te_3$. *Physical review letters* 103, 266801 (2009).
38. T Zhang et al. Experimental demonstration of topological surface states protected by time-reversal symmetry. *Physical Review Letters* 103, 266803 (2009).
39. P Sessi et al. Dual nature of magnetic dopants and competing trends in topological insulators. *Nature communications* 7, 12027 (2016).
40. C-Z Chang et al. High-precision realization of robust quantum anomalous hall effect in a hard ferromagnetic topological insulator. *Nature materials* 14, 473-477 (2015).

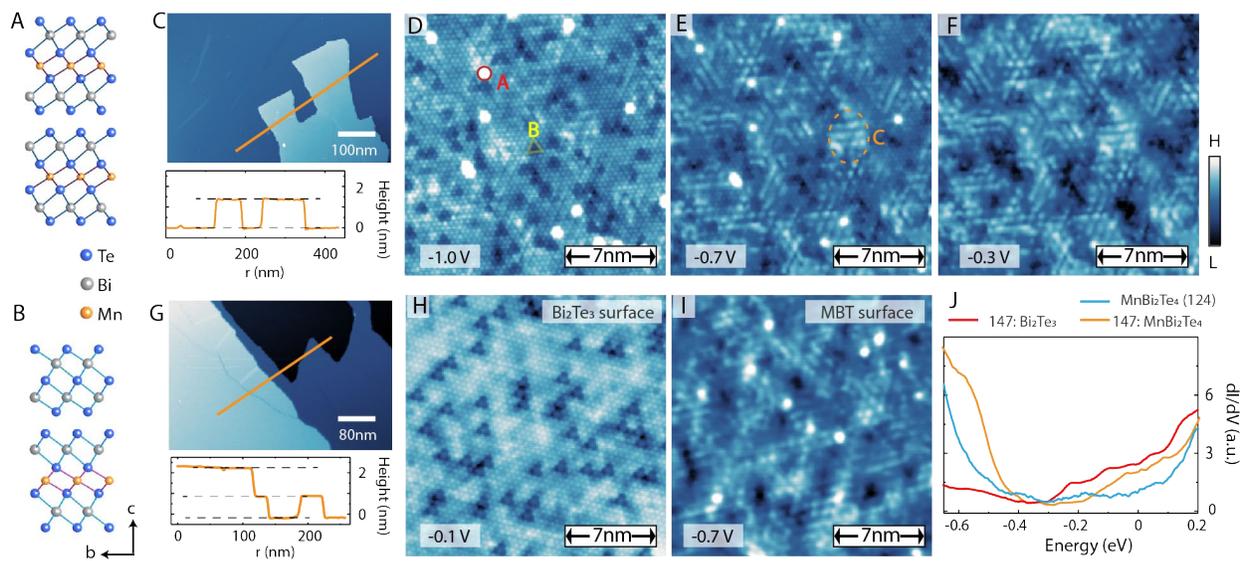

**Figure 1**

**Surface conditions of Mn124 and Mn147.** (A–B) Schematic of the crystal structures of Mn124 and 147. (C, G) Topographies showing step edges. (D-F) Atomic-resolution topographic images obtained at different bias voltages in Mn124. (H-I) Atomic-resolution topographic images obtained on different surfaces of Mn147. (J) Tunneling spectra taken on the three types of surfaces.

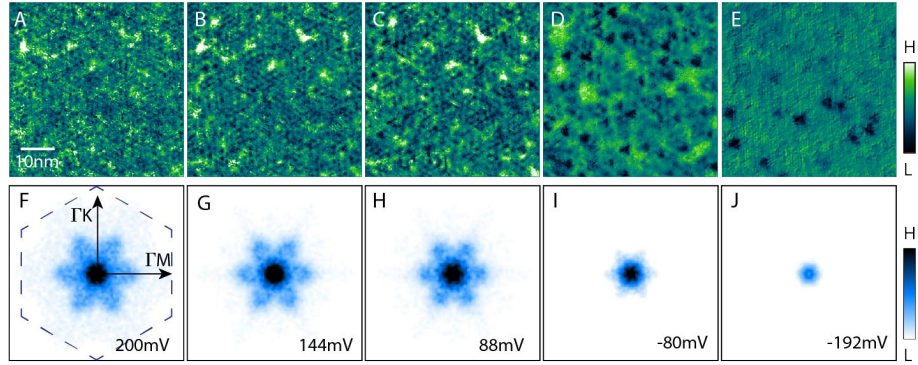

**Figure 2**

**Quasiparticle interference of Mn124.** (A-E) normalized dI/dV maps in a 50nm-square area with different energies. (F-J) Drift-corrected and symmetrized Fourier transforms of the dI/dV maps. The dashed hexagon denotes the surface Brillouin Zone.

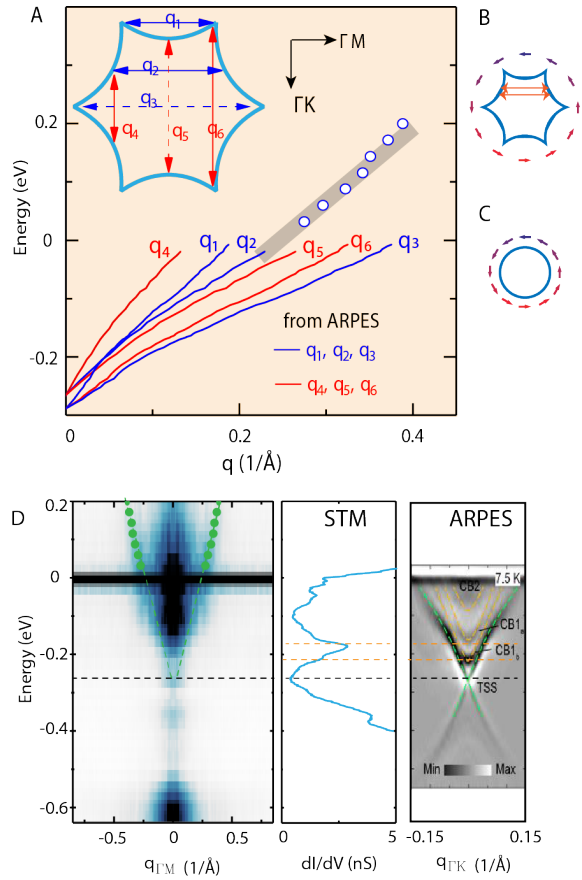

**Figure 3**

**Identification of the FT-QPI patterns.** (A) Dispersion of the six scattering channels calculated from ARPES data (22, 26) and the comparison with QPI data. Inset shows all six scattering channels with the warped constant energy contour (CEC). (B-C) Schematic of CECs of the surface state and the primary scattering vectors observed in QPI. (D) Intensity plot of the FFT linecut reveals the dispersion of surface state, which matches well with the single dI/dV spectrum and ARPES data (22).

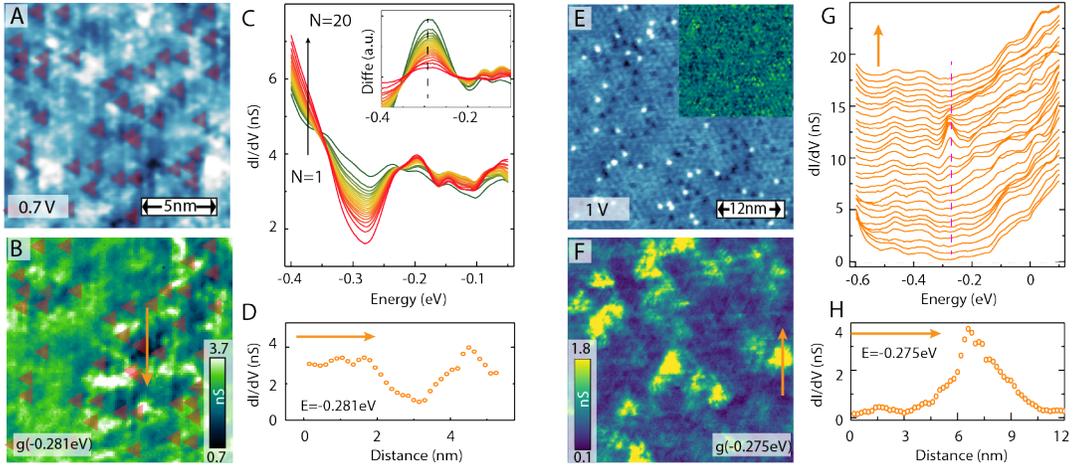

**Figure 4**

**Electronic inhomogeneity and impurity-induced states near Dirac energy in Mn124 and Mn147.** (A-B) Topographic image and the differential conductance obtained near the Dirac energy in Mn124. (C) Binned spectra show the electronic states generated near the Dirac point. Inset shows the difference between the spectra. (D) Spatial dependent of the resonance state. (E-F) Topographic image and the differential conductance obtained near Dirac energy in Mn147. Inset in E shows the real space QPI modulations far above the Dirac point. (G) Tunneling spectra taken along a line. Strong resonance states energetically located close to the Dirac point emerge. (H) Spatial evolution of the resonance state. The decay length is about a few nanometers.